\documentclass[3p]{elsarticle}

\usepackage{amsmath}

\newcommand{\alessio}{$^\dagger$}
\newcommand{\zimmer}{$^\ddagger$}
\newcommand{\tempdegree}{$^\circ\mathrm{C}$}
\newcommand{\smax}{$s_{\mathrm{max}}$}
\newcommand{\micron}{$\mu \mathrm{m}$}

\makeatletter
\def\@fnsymbol#1{\ensuremath{\ifcase#1\or \or \or
    \mathsection\or \mathparagraph\or \|\or **\or \dagger\dagger
    \or \ddagger\ddagger \else\@ctrerr\fi}}
\makeatletter

\usepackage{etoolbox}
\makeatletter
\patchcmd{\ps@pprintTitle}
  {Preprint submitted}
  {} 
  {}{}
\patchcmd{\ps@pprintTitle}
  {Elsevier}
  {} 
  {}{}
\patchcmd{\ps@pprintTitle}
  {to}
  {} 
  {}{}

\begin{document}

\begin{frontmatter}

\title{In-flight performance of the DAMPE silicon tracker}

\author[gva]{A. Tykhonov\corref{cor1}}
\cortext[cor1]{Corresponding author}
\ead{andrii.tykhonov@cern.ch}

\author[infn_pga]{G. Ambrosi}


\author[gva]{R. Asfandiyarov}

\author[gva]{P. Azzarello}

\author[mat_lec,infn_lec]{P. Bernardini}

\author[geo_pga,infn_pga]{B. Bertucci}

\author[geo_pga,infn_pga]{A. Bolognini\alessio\footnote{\alessio Now at RUAG Space, Zurich, Switzerland }} 

\author[gva]{F. Cadoux}










\author[mat_lec,infn_lec]{A.~D'Amone}

\author[mat_lec,infn_lec]{A.~De~Benedittis}

\author[gran_sasso,infn_gs]{I.~De~Mitri}

\author[mat_lec,infn_lec]{M.~Di~Santo}



\author[ihep]{Y. F. Dong}




\author[geo_pga,infn_pga]{M. Duranti}

\author[sassari,catania,asi]{D. D'Urso}

\author[ihep]{R. R. Fan}





\author[infn_bari,fis_bari]{P. Fusco}

\author[gva]{V. Gallo}


\author[ihep]{M. Gao}


\author[infn_bari]{F. Gargano}

\author[infn_pga,geo_pga]{S. Garrappa}

\author[ihep]{K. Gong}







\author[infn_pga]{M. Ionica}





\author[gva]{D. La Marra}

\author[infn_bari,fis_bari]{F. Loparco}







\author[mat_lec,infn_lec]{G. Marsella}

\author[infn_bari]{M.N. Mazziotta}




\author[ihep]{W. X. Peng}

\author[ihep]{R. Qiao}


\author[gva]{M. M. Salinas}









\author[infn_lec]{A. Surdo}




\author[geo_pga,infn_pga]{V. Vagelli}

\author[gva]{S. Vitillo}



\author[ihep]{H. Y. Wang}

\author[ihep]{J. Z. Wang}

\author[gran_sasso,infn_gs]{Z. M. Wang}

\author[ihep]{D. Wu}




\author[gva]{X. Wu}

\author[ihep]{F. Zhang}


\author[ihep]{J. Y. Zhang}

\author[ihep]{H. Zhao}






\author[gva]{S. Zimmer\zimmer\footnote{\zimmer Now at University of Innsbruck, Austria}}

\address[gva]{Department of Nuclear and Particle Physics, University of Geneva, CH-1211, Switzerland}
\address[infn_pga]{Istituto Nazionale di Fisica Nucleare Sezione di Perugia, I-06123 Perugia, Italy}
\address[mat_lec]{Universit\`a del Salento - Dipartimento di Matematica e Fisica "E. De Giorgi", I-73100, Lecce, Italy}
\address[infn_lec]{Istituto Nazionale di Fisica Nucleare (INFN) -- Sezione di Lecce , I-73100 , Lecce, Italy}
\address[geo_pga]{Dipartimento di Fisica e Geologia, Universit\`a degli Studi di Perugia, I-06123 Perugia, Italy}
\address[gran_sasso]{Gran Sasso Science Institute (GSSI), Via Iacobucci 2, I-67100, L'Aquila, Italy}
\address[infn_gs]{Istituto Nazionale di Fisica Nucleare (INFN) -- Laboratori Nazionali del Gran Sasso, I-67100, L'Aquila, Italy}
\address[ihep]{Institute of High Energy Physics, Chinese Academy of Sciences, YuquanLu 19B, Beijing 100049, China}

\address[sassari]{Chemistry and Pharmacy Department, Universit\`a degli Studi di Sassari, Sassari 07100, Italy}
\address[catania]{Istituto Nazionale di Fisica Nucleare (INFN) --  Laboratori Nazionali del Sud, Catania 95123, Italy}

\address[asi]{SSDC-ASI via del Politecnico snc, 00133 Roma, Italy}
\address[infn_bari]{Istituto Nazionale di Fisica Nucleare Sezione di Bari, I-70125, Bari, Italy}
\address[fis_bari]{Dipartimento di Fisica "M.Merlin" dell'Univerisit\`a e del Politecnico di Bari, I-70126, Bari, Italy}


\begin{abstract}
DAMPE (DArk Matter Particle Explorer) is a spaceborne high-energy cosmic ray and gamma-ray detector, successfully launched in December 2015. It is designed to probe astroparticle physics in the broad energy range from few GeV to 100 TeV. The scientific goals of DAMPE include the identification of possible signatures of Dark Matter annihilation or decay, the study of the origin and propagation mechanisms of cosmic-ray particles, and gamma-ray astronomy. DAMPE consists of four sub-detectors: a plastic scintillator strip detector, a Silicon--Tungsten tracKer--converter (STK), a BGO calorimeter and a neutron detector. The STK is composed of six double layers of single-sided silicon micro-strip detectors interleaved with three layers of tungsten  for photon conversions into electron--positron pairs. The STK is a crucial component of DAMPE, allowing to determine the direction of incoming photons, to reconstruct tracks of  cosmic rays and to estimate their absolute charge (Z). We present the in-flight performance of the STK based on two years of in-flight DAMPE data, which includes the noise behavior, signal response,  thermal and mechanical stability, alignment and position resolution.
\end{abstract}


\end{frontmatter}


\section{Introduction}

The DArk Matter Particle Explorer (DAMPE) is a high-energy astroparticle satellite, successfully launched on December 17, 2015 from the Jiuquan Satellite Launch Center in China~\cite{Chang:dampe}. It is designed to detect electrons and photons in the energy range from few GeV to 10~TeV, as well as protons and cosmic-ray ions in the energy range from 10~GeV to 100~TeV with excellent energy resolution and direction reconstruction precision~\cite{TheDAMPE:2017dtc}. The main scientific objectives of DAMPE are the identification of possible signatures of Dark Matter annihilation or decay, a better understanding of the origin and propagation mechanisms of cosmic rays through the measurement of their flux and composition, and gamma-ray astronomy. 

\begin{figure}[]
\begin{center}
\includegraphics[width=0.5\textwidth]{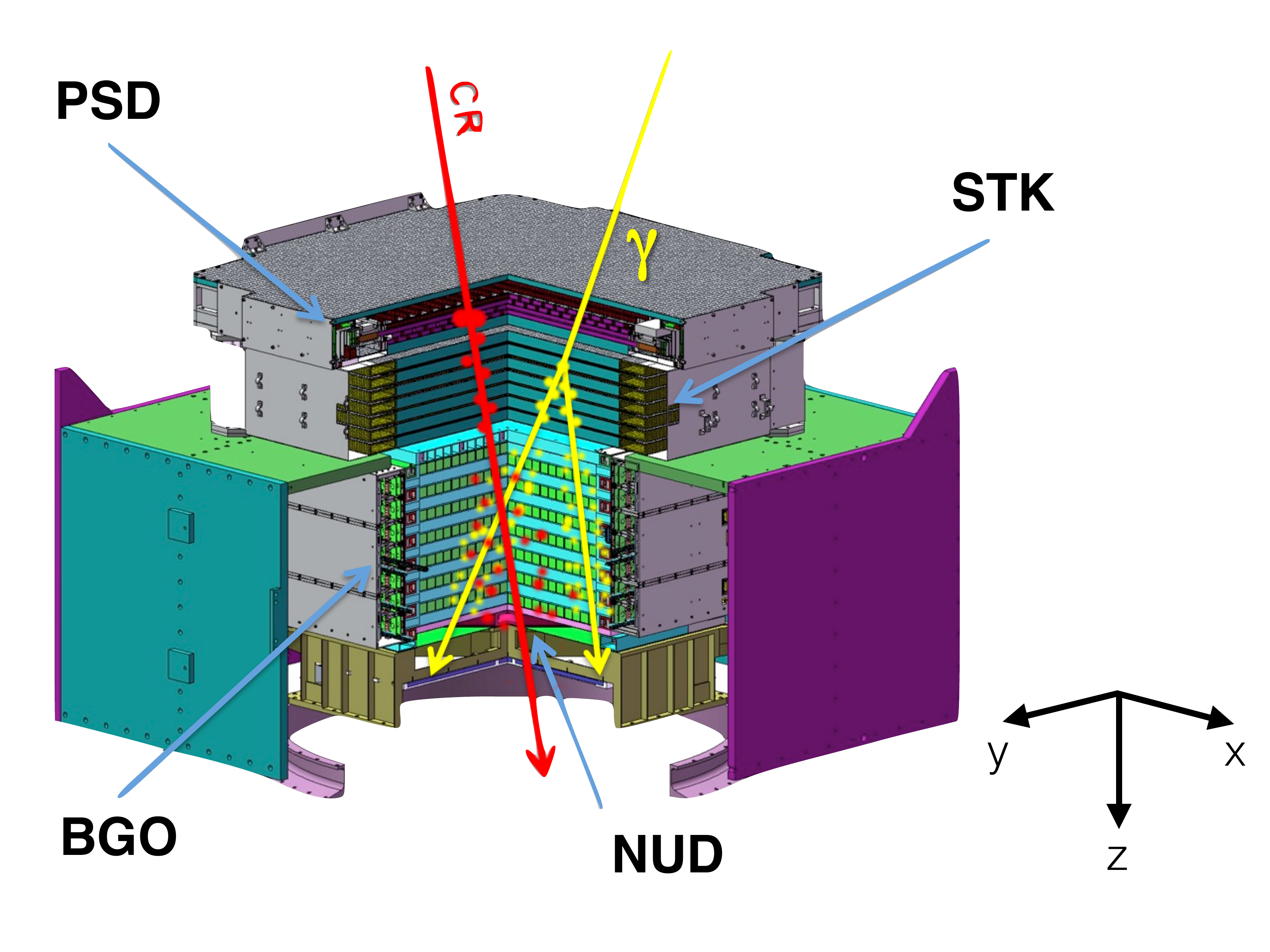}
\end{center}
\caption{Schematic view of the DAMPE satellite payload. It is composed of four sub-detectors: a plastic scintillator (PSD), a BGO calorimeter, a silicon-tungsten tracker (STK) and a Neutron Detector (NUD).}
\label{fig:dampe}
\end{figure}

The DAMPE  payload consists of four sub-detectors as shown in Figure~\ref{fig:dampe}. First, there is a plastic scintillator strip detector (PSD) which provides a veto signal for the photon detection and measures the absolute charge of cosmic-ray ions. The second sub-detector is a Silicon--Tungsten tracKer--converter (STK), which is described in detail in the next section. The third one is a Bismuth Germanium Oxide calorimeter (BGO) consisting of 14 layers in hodoscopic arrangement with a total thickness of about 32 radiation lengths, for a precise energy measurement and electron/proton discrimination. The last sub-detector is a boron-doped plastic scintillator which detects delayed neutrons originating from hadronic interactions at high energies (NUD), and enhances the electron/proton  separation.

The DAMPE satellite is operating on a sun-synchronous orbit at an altitude of about 500 km and inclination of 97$^{\circ}$, in a sky-survey mode 
permanently oriented to zenith. 
 Each orbit lasts about 95 minutes.

\section{The silicon--tungsten tracker--converter (STK)}

\begin{figure}[]
\begin{center}
\includegraphics[width=0.40\textwidth]{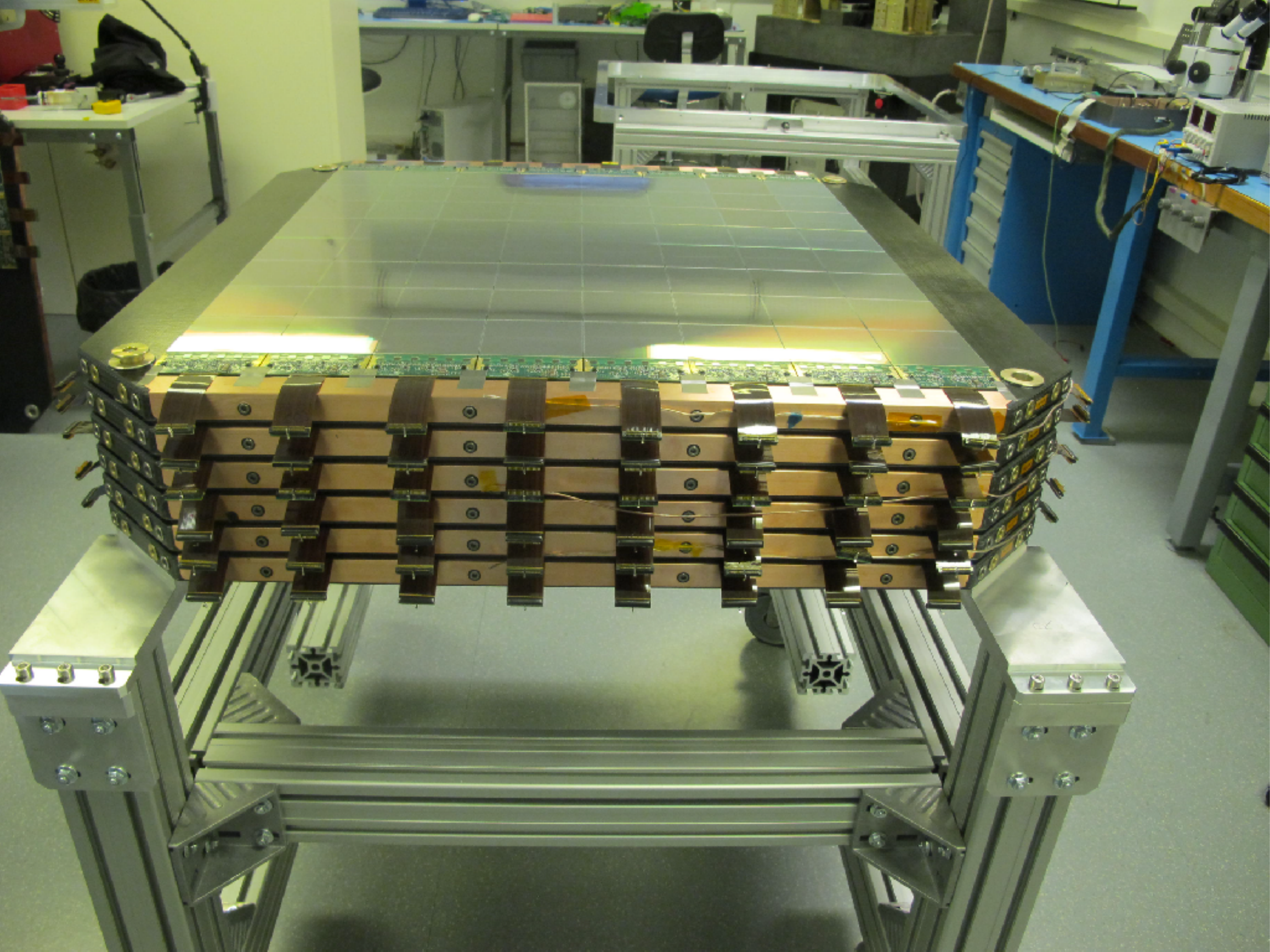}
\includegraphics[width=0.48\textwidth]{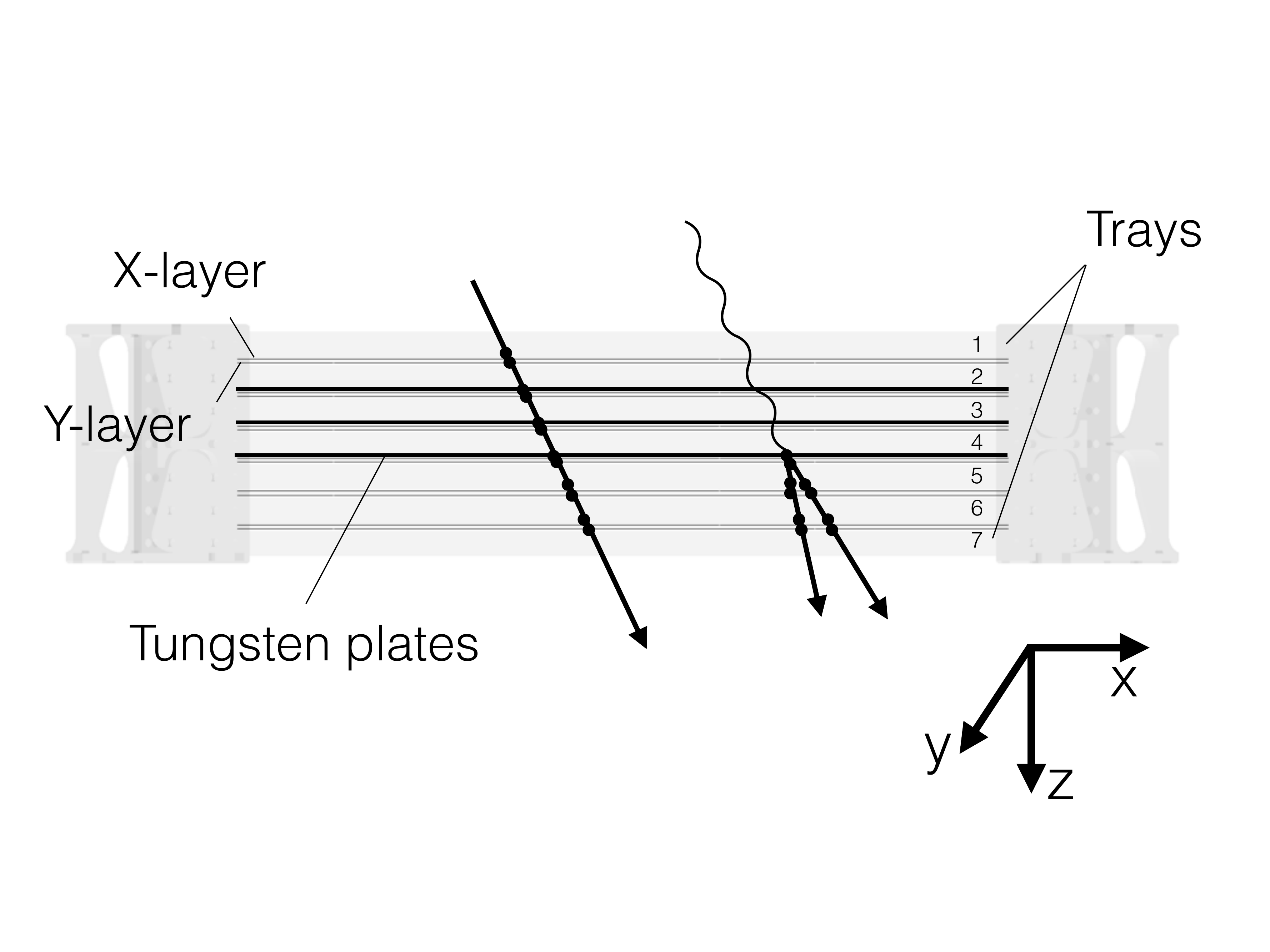}
\end{center}
\caption{Left: picture of the STK before the assembly of the last tray. The 64 silicon sensors are visible, together with the copper straps used for the heat transfer. Right: schematic view of the STK. The $x$  and $y$ layers are shown together with the tungsten converter planes, carbon-fiber support trays and aluminum corner feet structures. A passage of charged particle and a photon conversion in the STK is illustrated.}
\label{fig:stk}
\end{figure}

The STK is a crucial component of DAMPE aimed to identify the direction of incoming photons, reconstruct the  trajectories of charged particles and measure the absolute charge of cosmic-ray ions~\cite{Azzarello:2016trx,Gallo:2017tyu,Wu:2015fdz}. It consists of six double layers of single-sided silicon micro-strip detectors, mounted on seven support trays, providing the coordinate measurement in two orthogonal directions perpendicular to the satellite pointing direction, as shown in Figure~\ref{fig:stk}. It is interleaved with three layers of tungsten converters, 1~$\mathrm{mm}$ thick, inserted after the first, second and third tracking layers, to initiate the photon conversion into electron--positron pairs.

\begin{figure}[]
\begin{center}
\includegraphics[width=0.60\textwidth]{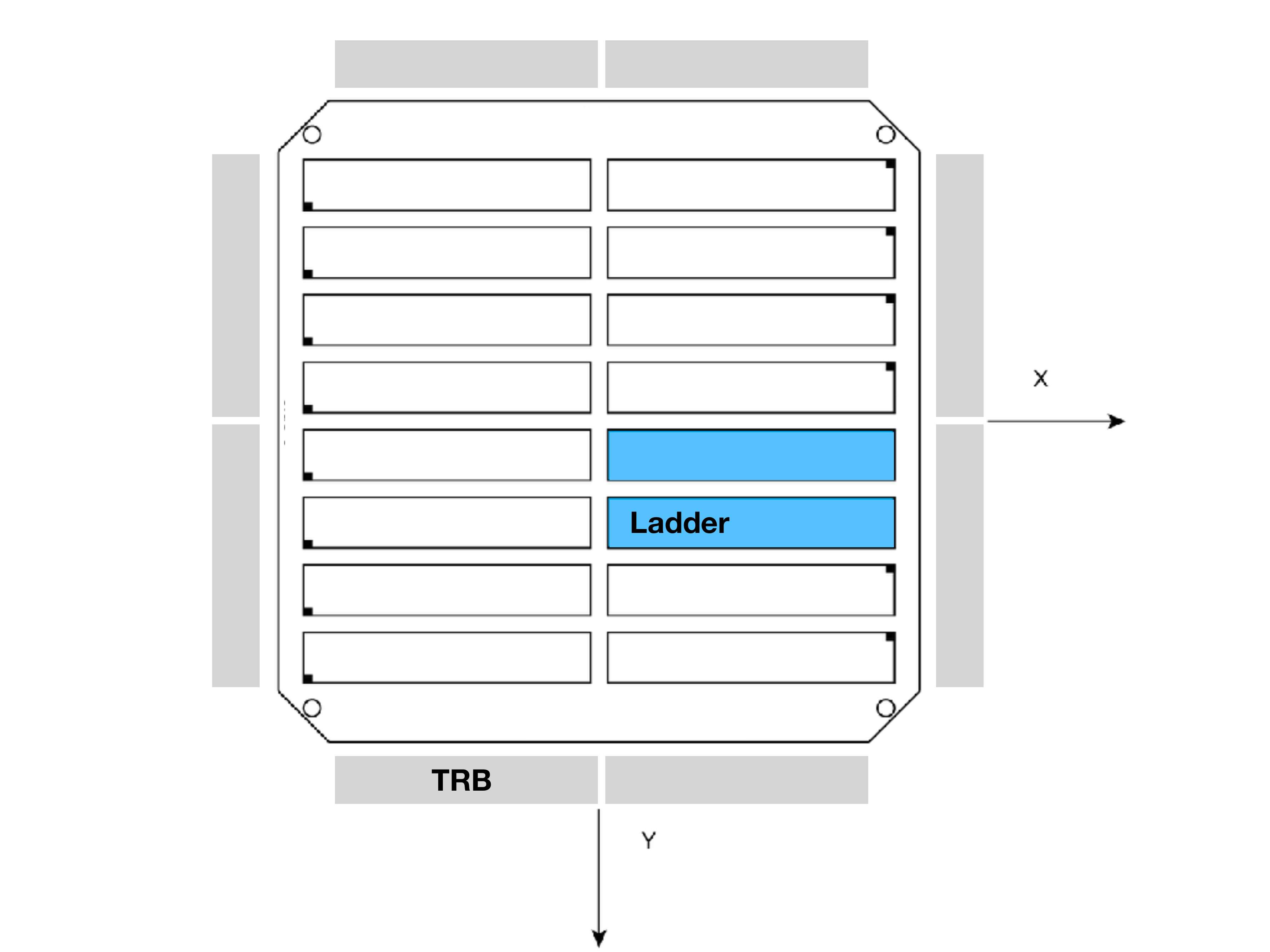}
\end{center}
\caption{Schematic view of the ladder assembly in the $y$ STK layer. The $x$ layer has a similar structure, with the ladders oriented along the $y$ axis. The TRBs are shown in pale gray.}
\label{fig:stk_assembly}
\end{figure}

Each tracking layer consists of 16 modules called ladders. Each ladder is made of 4 AC-coupled single-sided silicon micro-strip sensors daisy-chained via micro-wire bonds.  There are in total 192 ladders in the STK, read by 8 data acquisition Tracker Readout Boards (TRB)~\cite{Zhang:2016hth}. As shown in Figure~\ref{fig:stk_assembly}, the TRBs are installed on the sides of the STK (two boards per side)  where each TRB reads out 24 ladders.  

The silicon sensors are produced by Hamamatsu Photonics~\cite{hamamatsu} and are 320~$\mu \mathrm{m}$ 
 thick with dimensions of 9.5~$\mathrm{cm}$~$\times$~9.5~$\mathrm{cm}$. Each sensor is segmented into 768 strips with a pitch of 121~\micron. In order to limit the number of readout channels, density of readout electronics and power consumption while keeping a good performance in terms of position resolution, the readout is done for every other strip. This makes 384 readout channels per ladder, read by 6 VA140 ASICs produced by IDEAS~\cite{ideas}. The VA chips are well known components used also in space projects like AMS-01 and AMS-02~\cite{Ambrosi:1999sc}.  Owing to analog readout and charge sharing on the non-readout (floating) strips, the position resolution of the silicon sensors is better than 70~\micron~for most incidence angles~\cite{Azzarello:2016trx}.

The raw data of each STK readout channel is presented as a 12 bit ADC value, with a total of 73728 readout channels in the STK. Each ADC value can be considered as a sum of four components:
\begin{equation*}
\mathrm{ADC}_{ij} = \mathrm{ped}_i + \mathrm{cn}_{j} + \mathrm{noise}_{ij} + \mathrm{signal}_{ij} \\
\end{equation*}
where $i$ and $j$  is the channel and event number respectively. The first component is the channel \emph{pedestal value}. It is calculated by averaging the ADC values in $n$ (1024) consecutive events, collected with the internal-clock (calibration) trigger:
\begin{equation*}
\mathrm{ped}_i=\frac{1}{n}\sum_{j=1}^{n}\mathrm{ADC}_{ij}
\end{equation*}
The second component is the \emph{common noise}, calculated per-event by averaging the pedestal-subtracted ADC values of 64 channels belonging to the same VA chip:
\begin{equation*}
\mathrm{cn}_j=\frac{1}{64}\sum_{i=1}^{64}(\mathrm{ADC}_{ij}-\mathrm{ped}_i)
\end{equation*}
To avoid the signal contamination in the common noise evaluation, channels with $\mathrm{ADC}_{ij}-\mathrm{ped}_i>30$~ADC counts are excluded in the above computation.
The last two components are the channel \emph{intrinsic noise} and the \emph{signal} respectively. The average intrinsic noise, $\sigma_i$, can be calculated using the $n$ consecutive internal-trigger events, as follows:
\begin{equation*}
\sigma_i = \sqrt{\frac{1}{n}\sum_{j=1}^n(\mathrm{ADC}_{ij} - \mathrm{ped}_i -  \mathrm{cn}_{j}  )^2}
\end{equation*}
The $\sigma_{i}$ is simply referred to as \emph{noise} in the remainder of the paper.

The ADC values are processed on-board with the 16 FPGAs located in the TRBs (two FPGAs in each TRB) to  reduce the data flow from the satellite~\cite{Dong:2015qma}. The on-board data reduction algorithm performs simple signal clustering starting with the seeds with $S/N>3.5$ and adding up channels with the signal higher than 5 ADC counts. The average channel  noise is about 2.85 ADC counts, while the maximal probable value of the signal of a MIP particle hitting the readout (floating) strip is about 52 (27)  ADC counts.   A dedicated cluster reconstruction is then performed offline and the track reconstruction is done using a custom adaptation of Kalman filter algorithm~\cite{Tykhonov:2017uno}.

The STK layers are mechanically assembled on seven support trays composed of aluminum honeycomb sandwiched between Carbon Fiber Reinforcement Polymer (CFRP) sheets. The latter are 0.6~$\mathrm{mm}$ and 1.0~$\mathrm{mm}$ thick for layers with and without tungsten respectively. The trays form a light but rigid structure able to sustain vibrations and shocks during the rocket launch. In the second, third and fourth trays 1~$\mathrm{mm}$ tungsten converter plates are glued on the CFRP inside the tray. The converters are placed right before the tracking layers to ensure a high performance of photon vertex reconstruction. The trays were produced by Composite Design S\`{a}rl~\cite{composite}.  
 The layers including tungsten were X-ray scanned at CERN to check for optimal placement of the tungsten plates.

The construction of the STK Flight Model (FM) was completed in April 2015. The FM was then tested for one week with cosmic rays at the University of Geneva. In particular, a preliminary measurement of the alignment was performed with cosmic-ray muons~\cite{TheDAMPE:2017dtc}. Then it was delivered to China for the vibration and thermal vacuum tests, which were performed in Beijing in May 2015. It successfully passed the payload and satellite integration tests and showed excellent performance, with a total number of noisy channels (noise$>$5 ADC counts) less than 0.3\%~\cite{Wu:2015fdz}. 

The next sections will give an overview of the STK on-orbit performance, whose evaluation is based on the in-flight data collected by the DAMPE satellite during a two-year period, from January 2016 to December 2017.

\section{Temperature and noise behavior}

\begin{figure}[]
\begin{center}
\includegraphics[width=0.85\textwidth]{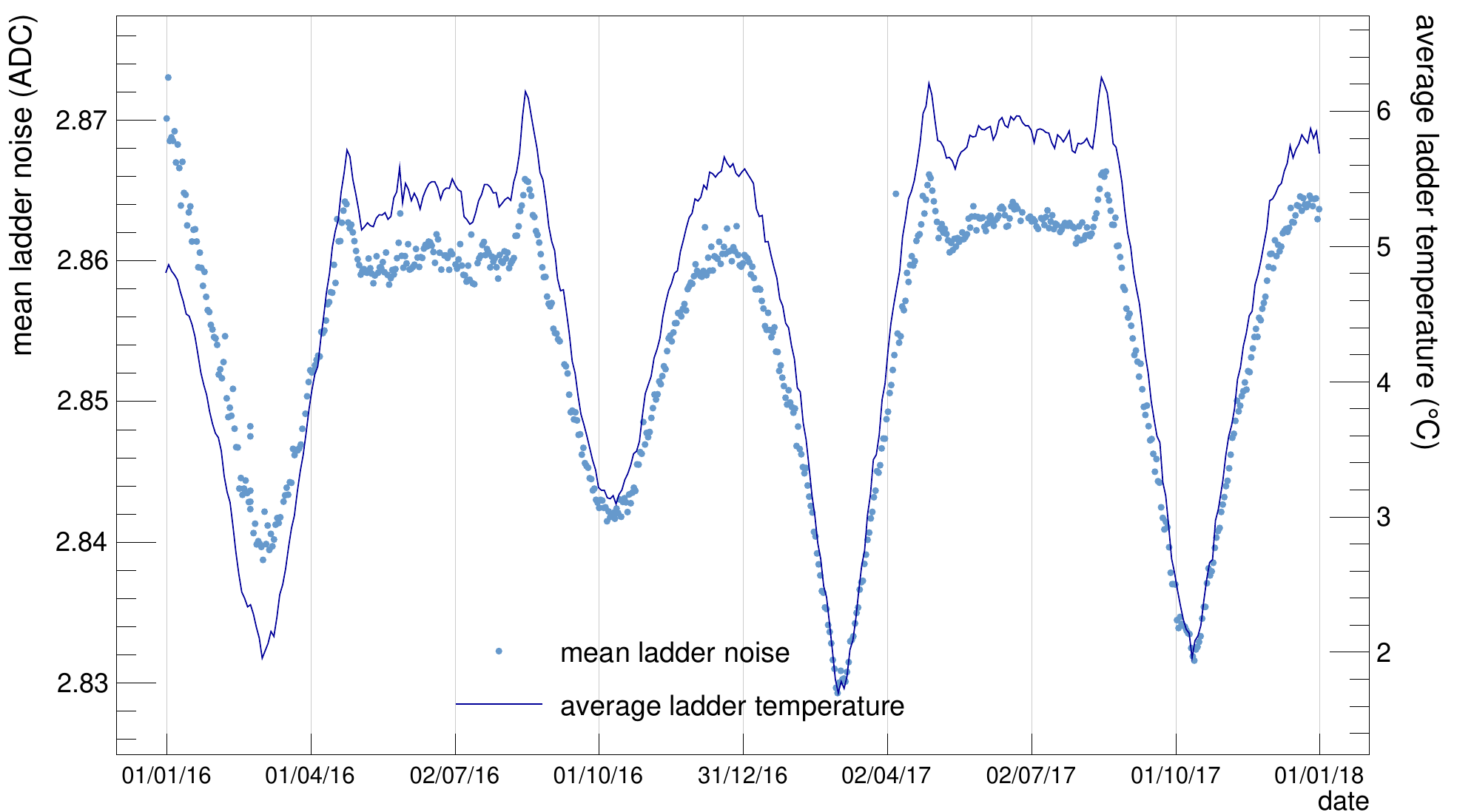}
\includegraphics[width=0.85\textwidth]{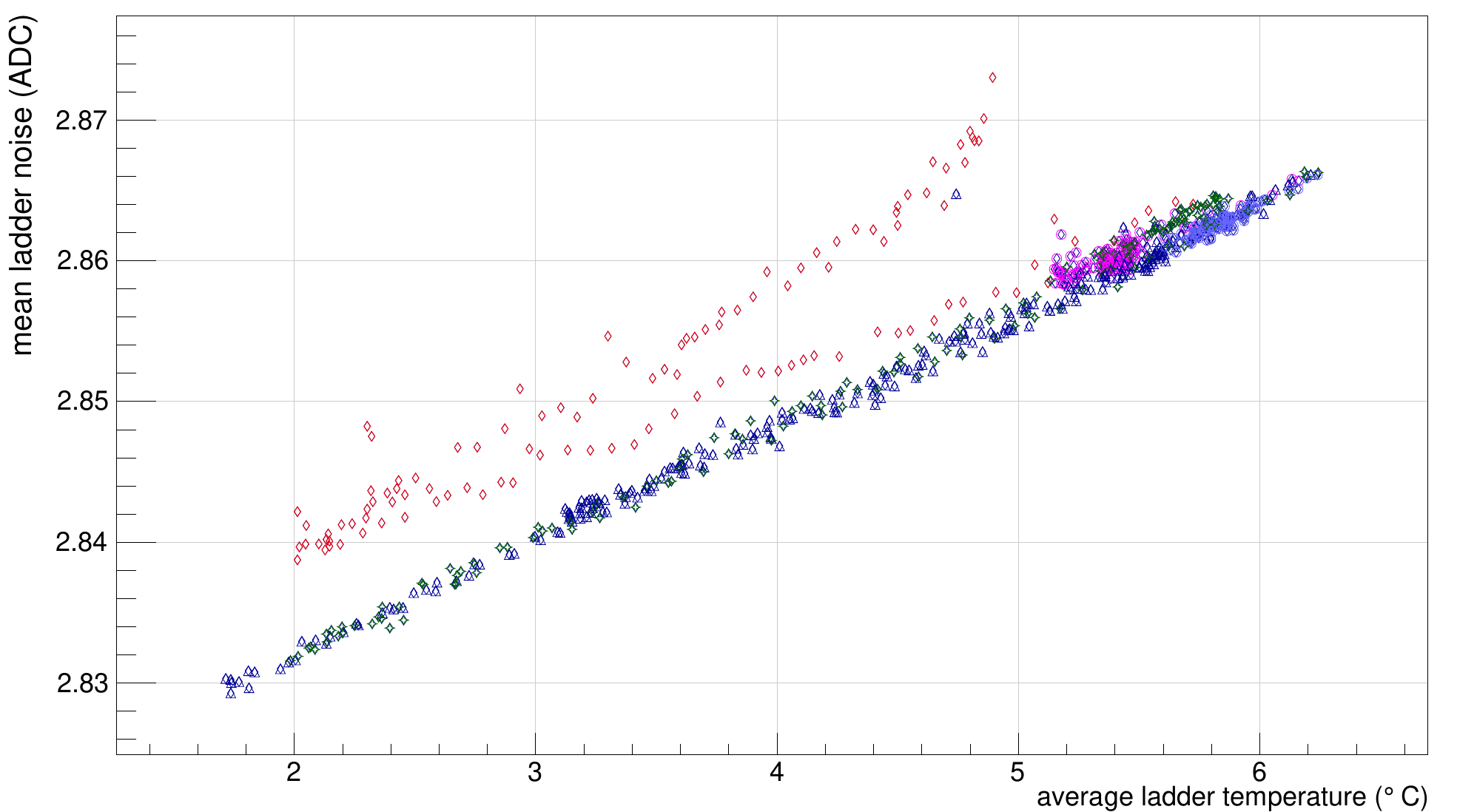}
\end{center}
\caption{Top: Evolution of the average ladder temperature (solid line) and the average STK channel noise (dotted line) from January 2016 to December 2017. Bottom: average ladder noise with respect to the average ladder temperature shown for different time periods: January--April 2016 (red diamonds), May--August 2016 (purple circles), September 2016 -- April 2017 (blue triangles), May--August 2017 (light blue circles) and September--December 2017 (green diamonds); a deviation of the noise-temperature correlation in the initial period is due to stabilization of the detector in space.} 
\label{fig:ladder_noise}
\end{figure}

\begin{figure}[]
\begin{center}
\includegraphics[width=0.85\textwidth]{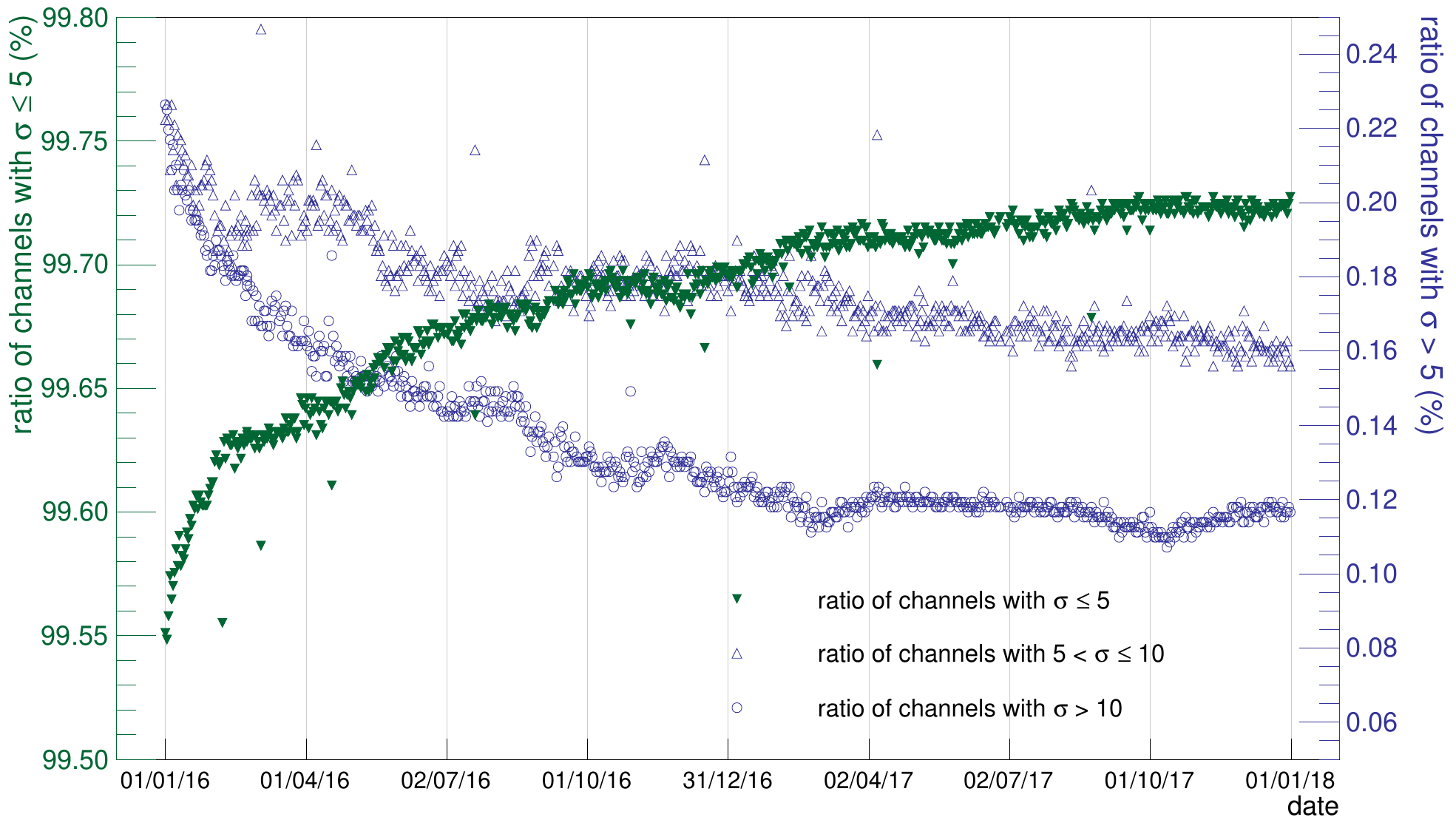}
\end{center}
\caption{Fraction of STK channels with noise below 5, between 5 and 10, and above 10 ADC counts for the same period.}
\label{fig:ladder_noise_2}
\end{figure}

The temperatures of the STK ladders are very stable, with a daily variation much lower than 1\tempdegree~(Figure~\ref{fig:ladder_noise}~top), as expected from the mechanical and thermal design. The small seasonal variation, amounting to maximum 4.4\tempdegree, is due to the orientation of the satellite orbit, moderated by the Earth's shadow from May to July. As a result the overall noise level of the STK is very stable: the average noise of all the 73728 channels has a maximum variation of 0.04 ADC counts, and is fully correlated with the temperature at a rate of about 0.01~ADC per~\tempdegree~(Figure~\ref{fig:ladder_noise}~bottom).  The excellent noise stability greatly simplified the in-flight operation of the STK, avoiding the complex and time consuming operation of updating on-board zero suppression thresholds from the ground when noise changes. The average noise is about 2.85~ADC, while the MIP signal for a particle hitting a readout (floating) strip is about 52~(27) ADC, achieving a signal/noise ratio of 18~(9). Another indicator of the excellent quality of the STK is the fraction of good channels (conservatively defined as noise lower than 5 ADC). From a level of 99.55\% at the beginning of data taking, it has improved over time because of the stabilization in space, reaching 99.72\% in December 2017 (Figure~\ref{fig:ladder_noise_2}).

\section{Signal gain calibration}

\begin{figure}[]
\begin{center}
\includegraphics[width=0.85\textwidth]{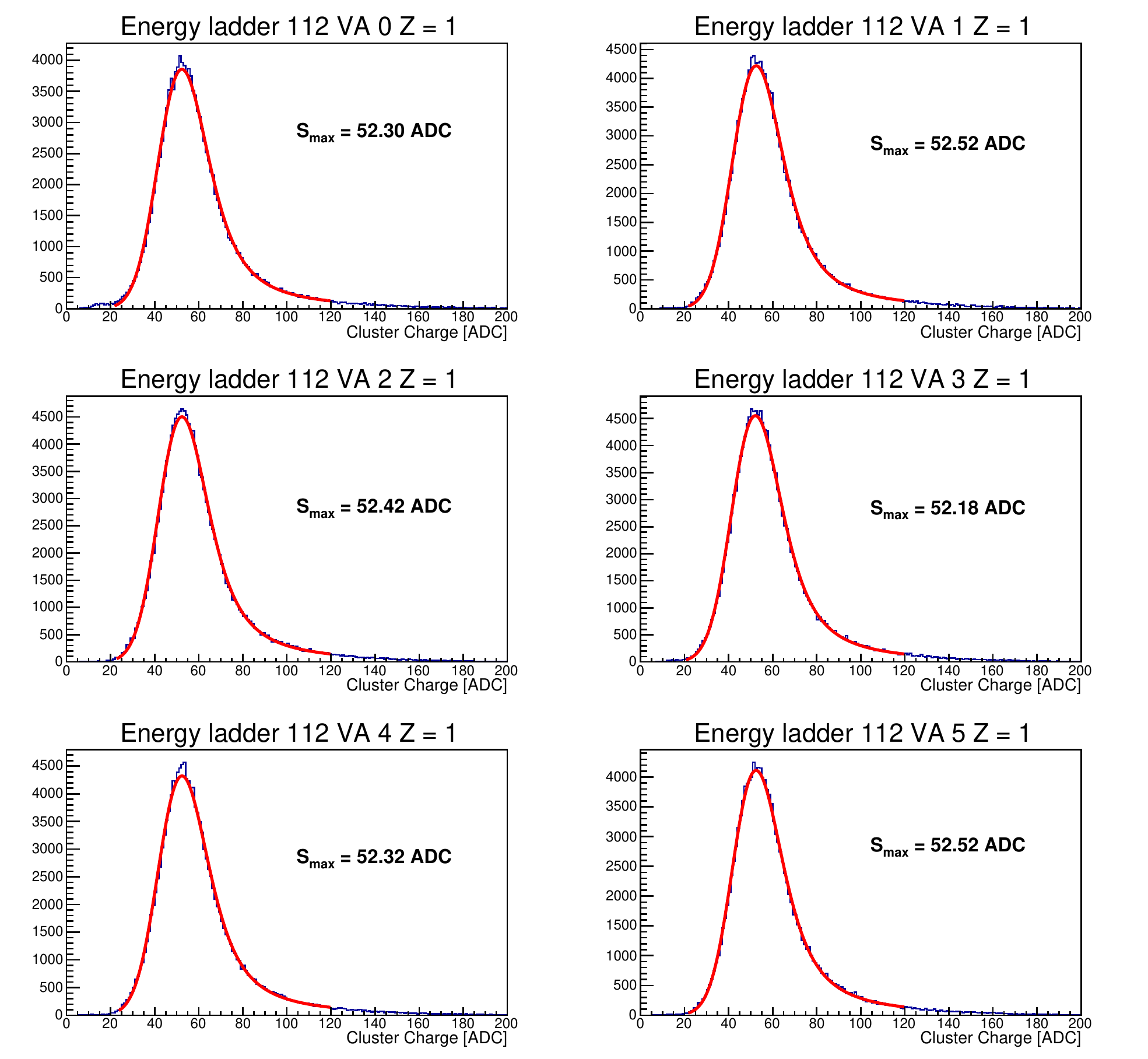}
\end{center}
\caption{Signal distribution of 6 VA140 chips belonging to the same ladder. Each histogram is fitted with a Landau distribution convoluted with a Gaussian function. The most probable value of the fit function (\smax) is shown for each VA. The distributions are based on the two-month in-flight data collected over a period November--December 2017.} 
\label{fig:va_signals}
\end{figure}

\begin{figure}[]
\begin{center}
\includegraphics[width=0.49\textwidth]{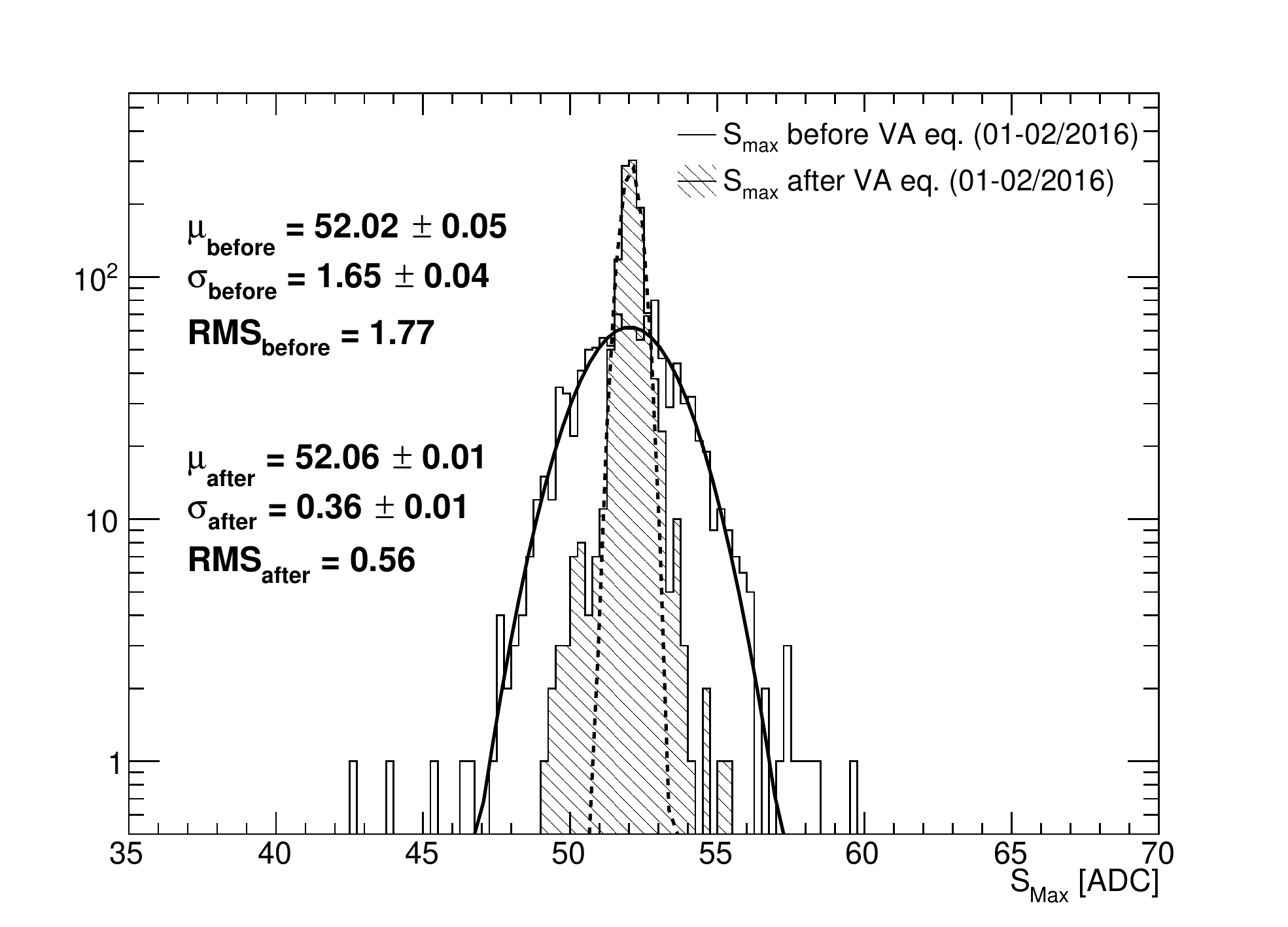}
\includegraphics[width=0.49\textwidth]{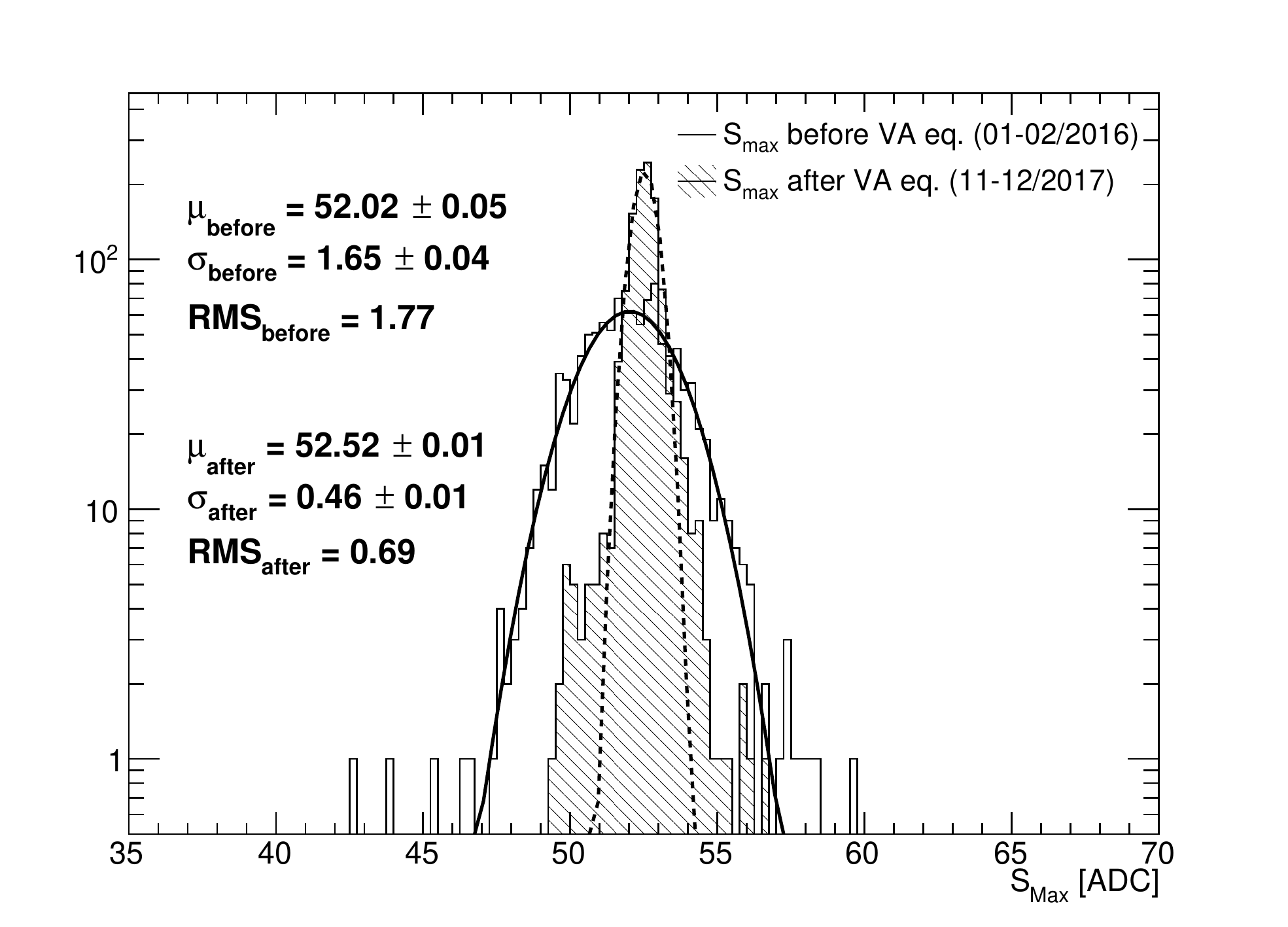}
\end{center}
\caption{The distribution of \smax for all VA140 chips of the STK before and after the gain calibration, fitted with a Gaussian function. Two different time periods are shown: January-February 2016 (left) and November-December 2017 (right).}
\label{fig:va_distributions}
\end{figure}

Each STK ladder is read out by 6 VA140 ASIC chips 
 dedicated to signal amplification and shaping~\cite{Zhang:2016hth}. This gives a total of 1152 chips which might have slightly different gain. Therefore, the gain calibration (equalization) is needed to ensure a uniform signal response for all chips, as discussed in~\cite{Gallo:2017tyu}. The calibration procedure is performed using the in-flight data with proton candidates collected over a two-month period, specifically January--February 2016. An example of signal distribution of 6 VA140 chips belonging to the same ladder is shown in Figure~\ref{fig:va_signals}. Particles at all incidence angles are considered and the signal is corrected for the particle path length in the silicon. Each histogram is fitted with a Landau distribution convoluted with a Gaussian function~\cite{Anticic:1996ja}, where the most probable value  of the fit function (\smax) corresponds to an optimum chip signal response. The distribution of~\smax~for all VA140 chips of the STK is shown as non-filled histograms in Figure~\ref{fig:va_distributions} and fitted with a Gaussian function. It has a standard deviation of about 1.7 ADC counts centered around a mean value of about 52 ADC counts. Therefore, each chip is calibrated to the mean value of this distribution.  The chip signal response after the calibration is shown as filled histograms in Figure~\ref{fig:va_distributions} for two different time periods.  As a result of the calibration, the width of the~\smax~distribution is less than 1 ADC count. The average~\smax~remains stable on-orbit with a variation of about 0.5 ADC counts, hence
  negligible for the particle hit and track reconstruction performance.
 
\section{Alignment and position resolution}

The position resolution of the STK silicon sensors is better than 70~\micron~\cite{Azzarello:2016trx}. At the same time,  the construction precision of the mechanical assembly of STK is about 100~\micron. To maximally profit from the position resolution capabilities of the silicon detectors,  alignment parameters are introduced to correct for shifts and rotations of each silicon sensor with respect to its position in the nominal instrument design. Alignment corrections are calculated based on the data and then used in the reconstruction procedure to correct the coordinates of particle hits. A detailed description of the STK alignment and track reconstruction procedure can be found in~\cite{Tykhonov:2017uno}. 

\begin{figure}[]
\begin{center}
\includegraphics[width=0.49\textwidth]{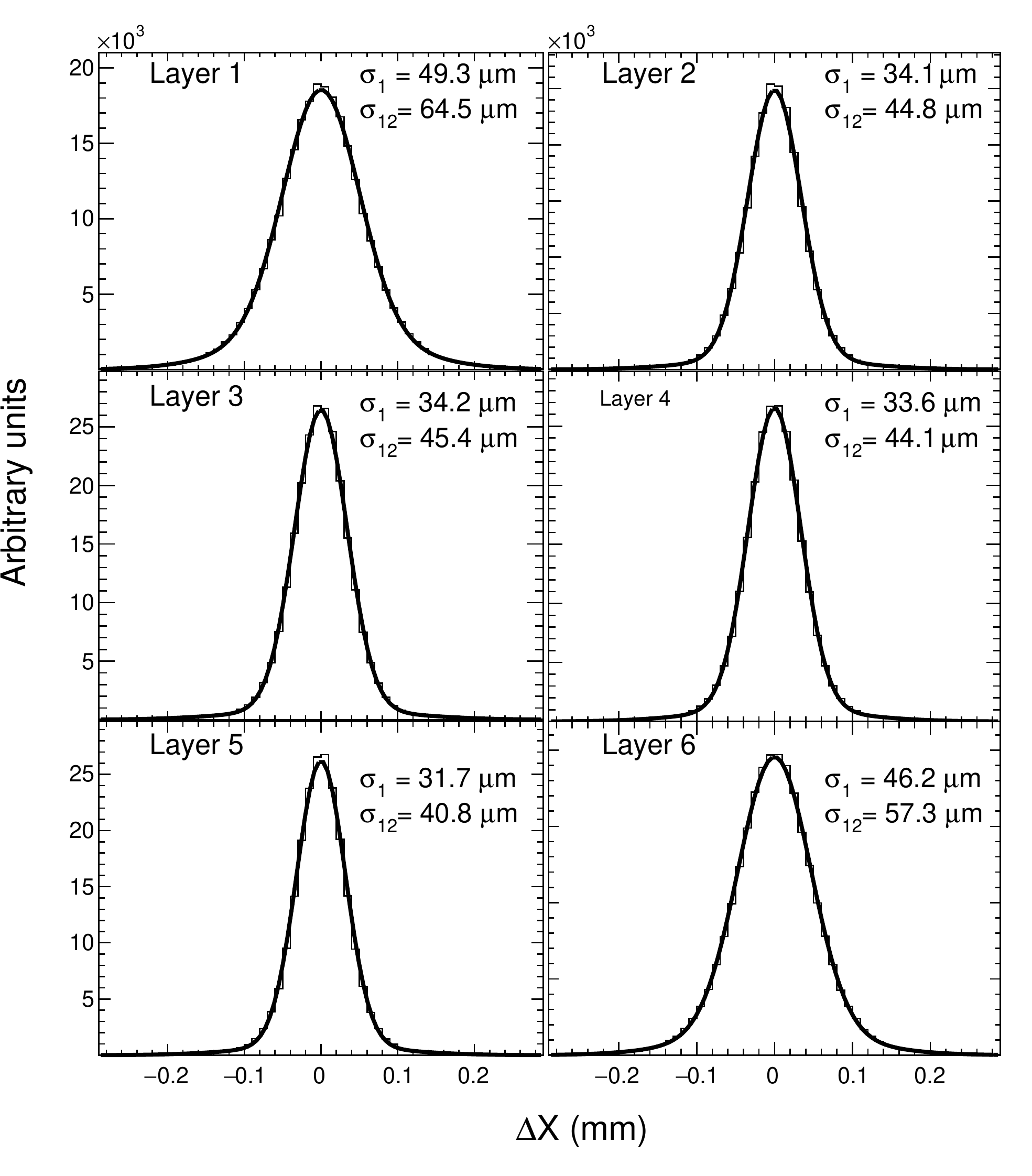}
\includegraphics[width=0.49\textwidth]{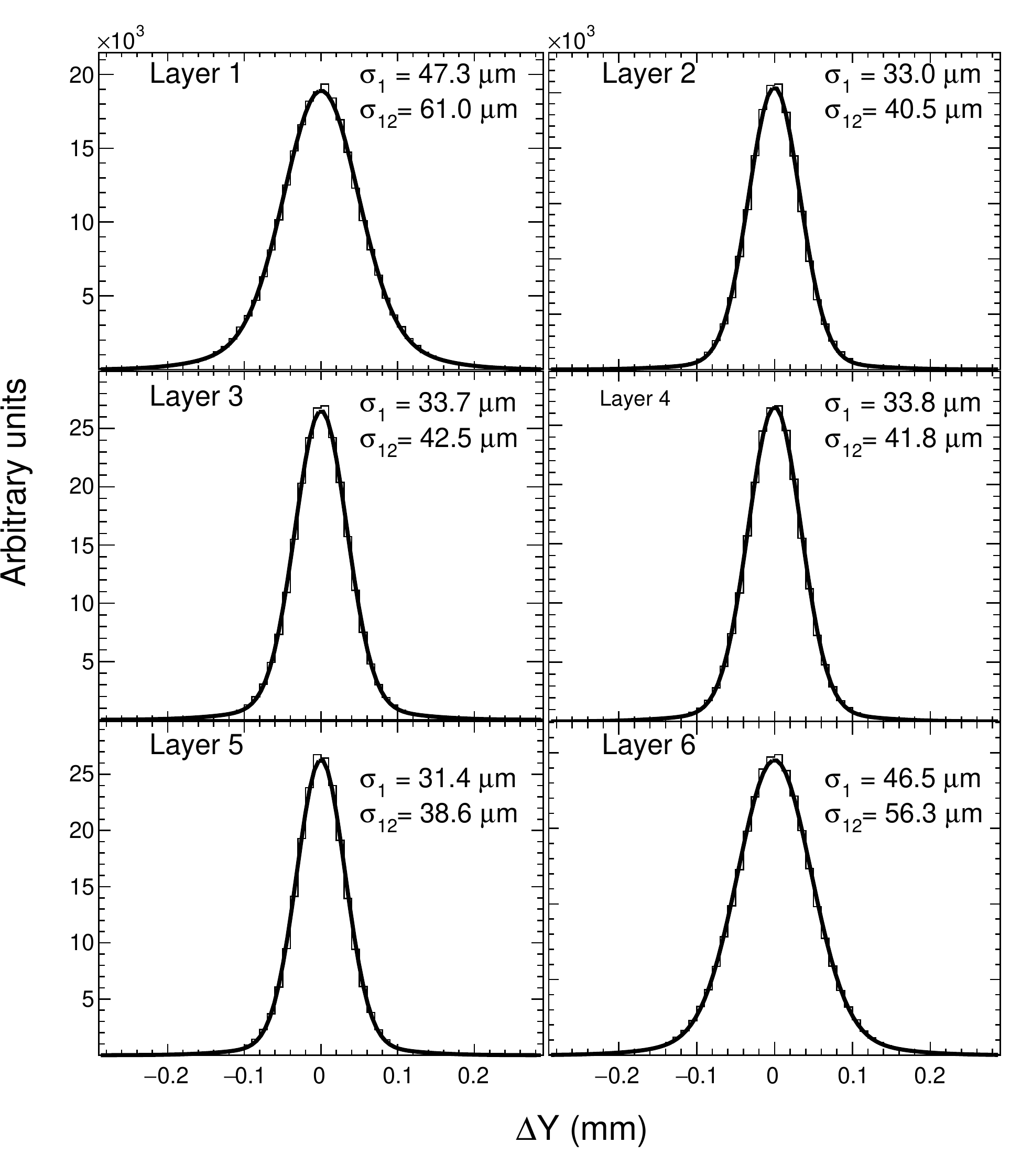}
\end{center}
\caption{Distribution of track-hit residuals for $x$ (left) and $y$ (right) layers of STK. Residuals are defined as measured minus projected hit positions, where the projection is obtained from a linear fit of the five remaining points of a track, without the point under study. All particle incidence angles are considered. The $\sigma_1$ and $\sigma_{12}$ parameters are the width of a narrower Gaussian and the width of area-weighted average of two Gaussians in the  fit of the residuals respectively.}
\label{fig:pos_resolution}
\end{figure}

As a quality criterion of the alignment we use the STK position resolution, which is estimated from the track-hit residual distributions. Residuals are defined as the difference between measured and projected coordinate of a hit belonging to a track, where projection is obtained from the five remaining hits of the track, without the point being tested. Figure~\ref{fig:pos_resolution} shows an example of residual distributions after the alignment, evaluated from five-day in-flight data collected in December 2017. In order to minimize the contribution of multiple scattering to the residuals, the difference between the measured and projected hit position in each of the remaining five points of a track is required to be less than 10~\micron. The residual distributions are fitted with a sum of two Gaussians, as follows:  
\begin{equation*}
N(x^{fit}-x^{hit}) = \frac{N_1}{\sqrt{2\pi}\sigma_1}e^{-\frac{(x^{fit}-x^{hit})^2}{2\sigma_1^2}} + \frac{N_2}{\sqrt{2\pi}\sigma_2}e^{-\frac{(x^{fit}-x^{hit})^2}{2\sigma_2^2}}
\label{eq:two_gaus}
\end{equation*}
\begin{equation*}
\sigma_{12} =\sqrt{\frac{N_1\sigma_1^2+N_1\sigma_2^2}{N_1+N_2}}
\end{equation*}
where $\sigma_{12}$ indicates the width of the area-weighted average of two Gaussians. The width of narrower Gaussian ($\sigma_1$) is used as an indicator of the stability of the alignment, as described below.

The mechanical structure of the STK stays stable on orbit. However, a small variation remains due to the humidity release process in the beginning of in-flight operation and due to temperature variations through the whole period on orbit. As discussed in~\cite{Tykhonov:2017uno}, in the initial period a contraction caused by the humidity release is expected to result in bending of the carbon-fiber support trays and therefore to contribute to the $z$-direction alignment. Then, the temperature variation causes the expansion/contraction of mechanical structure which also contributes to the $z$-alignment. The total off-plane ($z$) variation of the mechanical structure is less than 100~\micron~for all STK layers  throughout the in-flight epoch of DAMPE~\cite{Tykhonov:2017uno}.  The total in-plane ($x$ and $y$) variation was found below 1~\micron, which can be explained by the fact that the STK trays are mechanically fixed by four aluminum corner feet and four aluminum frames holding the TRBs, which prevent the relative shift and rotation of the planes in $x$/$y$ direction.

\begin{figure}[]
\begin{center}
\includegraphics[width=0.49\textwidth]{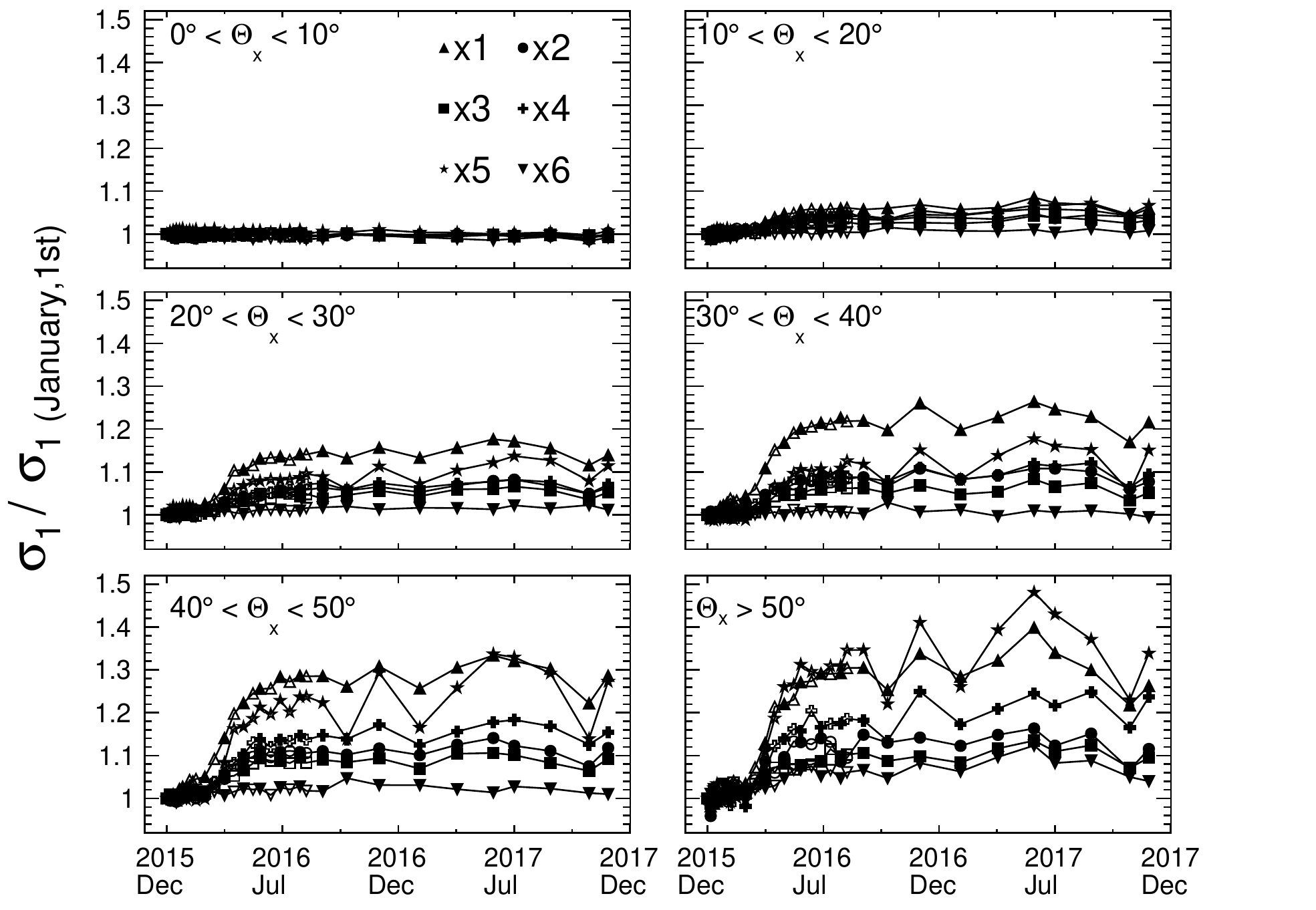}
\includegraphics[width=0.49\textwidth]{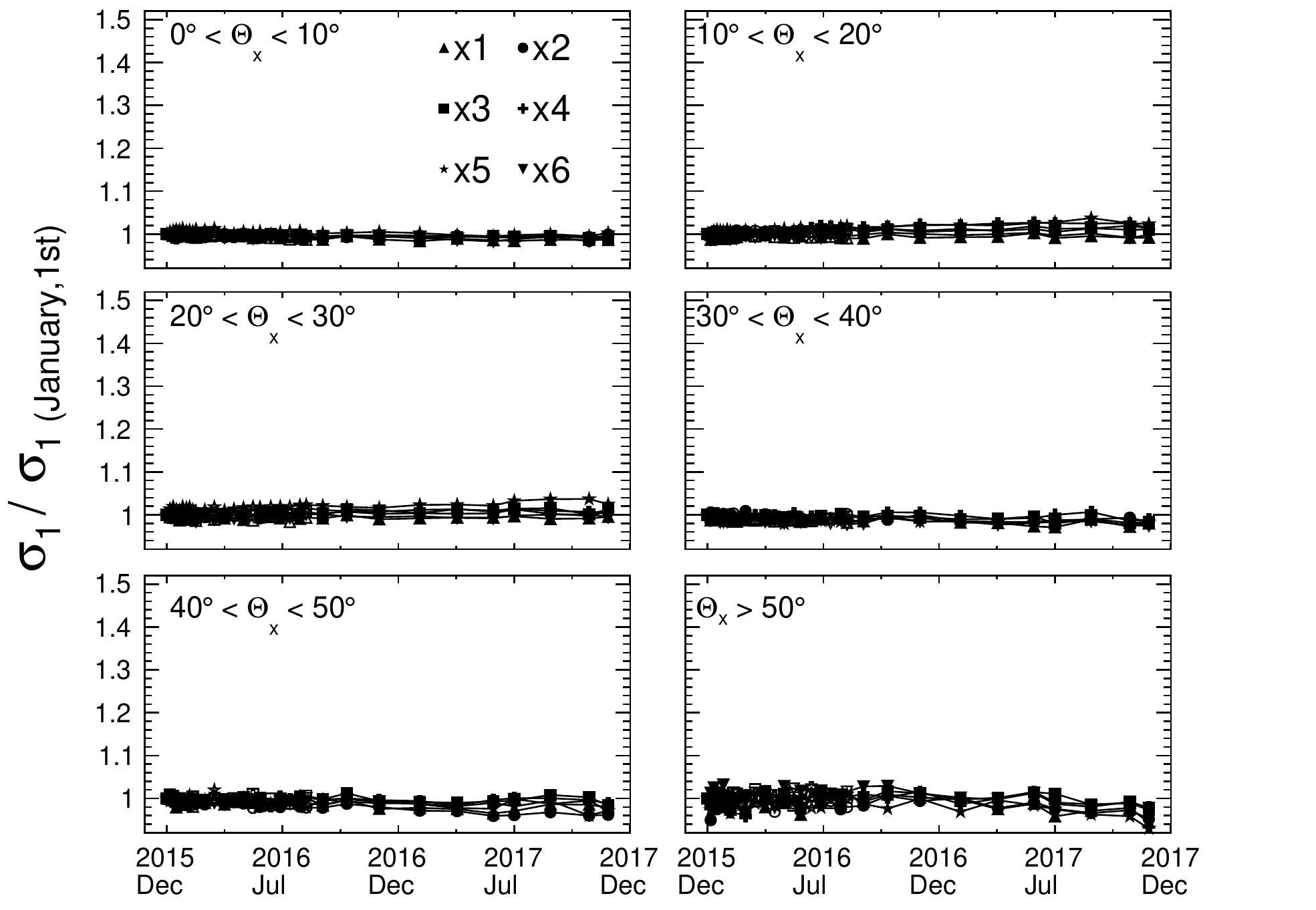}
\end{center}
\caption{Variation of $\sigma_1$ as a function of time for different $x$ layers of STK and different intervals of particle incidence angle, $\theta_x$. The cases of fixed (left) and time-dependent (right) alignment are shown. The $y$ layers show a similar behavior. }
\label{fig:pos_evolution}
\end{figure}

Given the good position resolution of the STK,  a regular re-calibration of the alignment has to be performed to account for the mechanical structure variations and to ensure the best position resolution. The alignment parameters are therefore updated every two weeks. 
 Figure~\ref{fig:pos_evolution} shows the variation of the $\sigma_1$ as a function of time. Two cases are shown to illustrate the impact of the alignment re-calibration: the single alignment (performed in January 2016)  and the time-dependent alignment. Thanks to the bi-weekly updates of the alignment parameters, the position resolution remains stable, the variation of $\sigma_1$ is below 4\% for all STK layers and particle incidence angles~\cite{Tykhonov:2017uno}. 

\section{Conclusions}

The Silicon--Tungsten tracKer--converter (STK) is a crucial sub-detector of the DAMPE mission, allowing to reconstruct  trajectories of charged particles, identify the absolute charge of cosmic-ray ions and pinpoint the direction of incoming gamma-rays. It is based on a well established technology of single-sided silicon micro-strip detectors with analog readout. After the launch the tracker has shown excellent performance. The total number of noisy channels is less than 0.3\% and steadily decreases with time. 
  The variation of channel noise is fully correlated with temperature at a very low rate of 0.01 ADC per~$^{\circ}$C. The total noise variation is below 0.04 ADC counts, which allows to avoid the computationally expensive threshold update procedure for the on-board data reduction.  Thanks to the gain equalization procedure, the signal gain is uniform for all the STK readout chips and stays stable with time, with a total variation less than 1 ADC count. The STK position resolution is better than 70~\micron~for all particle incidence angles and, owing to the regular updates of the alignment parameters, it remains stable with time, with a total variation less than 4\%.

\section*{Acknowledgments}

The DAMPE mission is funded by the strategic priority science and technology projects in space science of Chinese Academy of Sciences (\mbox{No.~ XDA04040000} and~\mbox{No.~XDA04040400}). In Europe the experiment is supported by the Swiss National Science Foundation under grant~\mbox{No.~ 200020\_175806} and the National Institute for Nuclear Physics (INFN), Italy. The authors wish to express their gratitude to the generosity of CERN for providing beam time allocation and technical assistance at the PS and SPS beam lines, as well as general logistics support.

\bibliography{mybibfile}

\end{document}